*Research Article*

# Temperature Dependence of Electrical Characteristics of Carbon Nanotube Field-Effect Transistors: A Quantum Simulation Study

**Ali Naderi,[1] S. Mohammad Noorbakhsh,[2] and Hossein Elahipanah[2]**

[1] *Electrical Engineering Department, Semnan University, Semnan, Iran*
[2] *Department of Electrical Engineering, Boroujen Branch, Islamic Azad University, Boroujen, Iran*

Correspondence should be addressed to Hossein Elahipanah, elahipanah@ieee.org

By developing a two-dimensional (2D) full quantum simulation, the attributes of carbon nanotube field-effect transistors (CNTFETs) in different temperatures have been comprehensively investigated. Simulations have been performed by employing the self-consistent solution of 2D Poisson-Schrödinger equations within the nonequilibrium Green's function (NEGF) formalism. Principal characteristics of CNTFETs such as current capability, drain conductance, transconductance, and subthreshold swing (SS) have been investigated. Simulation results present that as temperature raises from 250 to 500 K, the drain conductance and on-current of the CNTFET improved; meanwhile the on-/off-current ratio deteriorated due to faster growth in off-current. Also the effects of temperature on short channel effects (SCEs) such as drain-induced barrier lowering (DIBL) and threshold voltage roll-off have been studied. Results show that the subthreshold swing and DIBL parameters are almost linearly correlated, so the degradation of these parameters has the same origin and can be perfectly influenced by the temperature.

## 1. Introduction

Among many materials that have been proposed to supplement and, in the long run, possibly succeed silicon as a basis for nanoelectronics, carbon nanotubes (CNTs) have attracted the most attention due to their extraordinary electronic and optical properties [1–3]. In addition to the efforts to develop new electronic devices, direct bandgap one-dimensional (1D) nanostructures are attracting attention because of the desire to base both electronic and optoelectronic technologies on the same material [1]. Among the different 1-D materials, single-walled CNTs [2] have desirable and distinctive device properties. In particular, electronic transport properties of single-wall CNTs have attracted considerable experimental and theoretical interest [4–10]. For example, CNT field effect transistors (CNTFETs) have generated considerable interest in the past few years because of their quasi-ideal electronic properties and have recently reached a high level of performance [11–14]. Transistor devices made of semiconductor single-wall CNT [2] can be considered as simple silicon MOS field-effect transistors with the silicon material replaced by the carbon nanotube structures. These devices are one of the current leading technologies to replace MOSFETs [15–17].

The MOSFET device characteristics and circuit behavior that changes with the variation in temperature were reported in many papers [18–23]. It is expected that the temperature plays key role in the CNTFET performance and characteristics. Hence, the CNTFET performance has to be predicted in different temperatures. According to the best of our knowledge, no comprehensive investigation on CNTFET device has been reported that includes the effects of change in temperature. Therefore, in this paper by varying temperature the attributes of these devices have been investigated. These attributes of CNTFETs in different temperatures have been investigated using two-dimensional (2D) quantum simulation. The simulations have been done by the self-consistent solution of 2D Poisson Schrödinger formalism [24]. The effects of varying temperature are investigated in terms of on-current, leakage off-current, on-/off-current ratio, transconductance characteristics, drain conductance, subthreshold swing (SS), threshold voltage roll-off, and drain-induced barrier lowering (DIBL).

## 2. CNTFET Structure and Simulation Method

A schematic cross-sectional view of the simulated cylindrical CNTFET is shown in Figure 1. The device has a zigzag

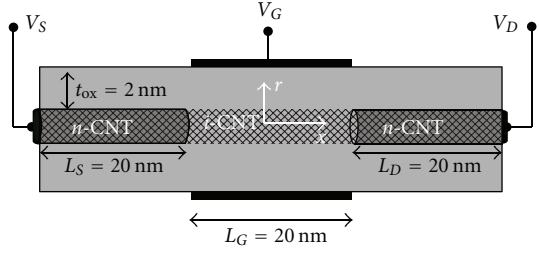

Figure 1: Schematic cross-sectional view of a coaxial carbon nanotube field-effect transistor (CNTFET) with cylindrical gate.

(13, 0) CNT structure with approximately 0.5 nm radius which is embedded in cylindrical gate insulator of $HfO_2$ with the thickness and dielectric constant ($k$) of 2 nm and 16, respectively [24]. The length and doping concentration of source and drain regions are 20 nm and $10^9$ nm$^{-1}$, respectively. The channel is intrinsic and its length is 20 nm. There is no overlap between source (drain) and gate regions.

In order to simulate the device characteristics, a nonequilibrium Green's function (NEGF) formalism has been used and the Schrödinger and Poisson equations have self-consistently been solved. The band structure of CNT has been calculated by the tight-binding method. The NEGF method provides a sound approach for the simulation of the nanoscale system out of equilibrium. Poisson equation simulates gate control on channel and transport equation simulates charge transfer between source and drain. The Poisson equation is solved to obtain the electrostatic potential in the nanotube channel. By solving the Schrödinger equation within the NEGF method, the density of states and the charge of the surface of the CNT can be obtained. By using the calculated charge and solving the Poisson equation the new electrostatic potential is developed. The iteration between Poisson and Schrödinger equations continues until the self-consistency is achieved.

Because of symmetric properties, it is convenient to solve Poisson's equations in cylindrical coordinates. Since the potential and charge are invariant around the nanotube, the Poisson equation is essentially a 2D problem along the tube ($z$-direction) and the radial direction ($r$-direction) as Poisson equation is written as [25, 26]

$$\frac{\partial^2 u_j(r,z)}{\partial r^2} + \frac{1}{r}\frac{\partial u_j(r,z)}{\partial r} + \frac{\partial^2 u_j(r,z)}{\partial z^2} = \frac{-q}{\varepsilon}\rho(r,z), \quad (1)$$

where $u_j(r,z)$ is the electrostatic potential, $\varepsilon$ is the dielectric constant, and $\rho(r,z_j)$ is the net charge density distribution which includes the dopant density as well. The net charge distribution $\rho(r,z_j)$ is given by

$$\rho(r=r_{CNT},z_j) = p(z_j) - n(z_j) + N_D^+ - N_A^-,$$
$$\rho(r \neq r_{CNT},z) = 0, \quad (2)$$

where $r_{CNT}$ is CNT radius, $N_D^+$ and $N_A^-$ are the ionized donor and acceptor concentrations, respectively. Then the computed electrostatic potential is used as input for the Schrödinger equation that is solved by using NEGF formalism. The retarded Green's function is computed by the following equation [5]:

$$G_q(E) = \left[(E + i\eta^+)I - H - \sum_S - \sum_D\right]^{-1}, \quad (3)$$

where $\sum_S$ and $\sum_D$ are the self-energies of the source and drain respectively, $\eta^+$ is an infinitesimal positive value, $E$ is the energy, $I$ is the identity matrix, and $H$ is the Hamiltonian of the CNT. The device Hamiltonian used in this paper is based on the atomistic nearest neighbor $p_z$-orbital tight binding approximation [27]. The cylindrical geometry of the device ensures symmetry in the angular direction, thus drastically simplifying the mode-space treatment of electron transport [28]. For an $(n,0)$ zigzag CNT with quantum number $q$, the Hamiltonian matrix for the subband is given by.

$$H = \begin{bmatrix} U_1 & b_{2q} & & & \\ b_{2q} & U_2 & t & & 0 \\ & t & U_3 & b_{2q} & \\ & & & \vdots & \\ & & & & b_{2q} & U_N \end{bmatrix}_{N \times N} \quad (4)$$

in which $b_{2q} = 2t\cos(\pi q/n)$, $t = 3$ ev is the nearest neighbor hoping parameter, and $N$ is the total number of carbon rings along the device. Here, the diagonal element $U_j$ corresponds to the on-site electrostatic potential along the tube surface obtained by solving the Poisson equation.

It is considered that a self-energy for semi-infinite leads as boundary conditions, and hence in these conditions, the CNT is connected to infinitely long CNTs at its ends. The source self-energy function ($\sum_S$) has all its entries zero except for the (1,1) element [28]:

$$\sum_S(1,1) = \frac{(E-U_1)^2 + t^2 + b_{2q}^2}{2(E-U_1)}$$
$$\pm \frac{\sqrt{\left[(E-U_1)^2 + t^2 + b_{2q}^2\right]^2 - 4(E-U_1)^2 t^2}}{2(E-U_1)}. \quad (5)$$

Similarly, $\sum_D$ has only its $(N,N)$ element nonzero, and it is given by an equation similar to above equation with $U_1$ replaced by $U_N$. After solving the Poisson equation and obtaining the electrostatic potential in the nanotube channel, this potential is used as input of transport equation. Self-consistency is achieved by iteration between Poisson and transport equation. The current is calculated by

$$I = \frac{2q}{h}\int T(E)[F(E-E_{FS}) - F(E-E_{FD})]dE. \quad (6)$$

This equation is Landaur-Buttiker formula. In this formula $T(E)$ is transmission coefficient, $E_{FS}$ and $E_{FD}$ are source and drain Fermi level, respectively, $q$ is the electron charge, and $h$ is Planck constant. $T(E)$ is calculated from the following equation:

$$T(E) = \text{trace}(\Gamma_s G \Gamma_D G^+), \quad (7)$$

where $G$ is Green's function, $\Gamma_{S(D)}$ is the energy level broadening due to source (drain) contact and is calculated from below equation:

$$\Gamma_{S(D)} = i\left(\sum_{S(D)} - \sum_{S(D)}^{+}\right) \quad (8)$$

In this paper the results are obtained from this simulation method.

## 3. Results and Discussion

In order to demonstrate the accuracy of quantum model, a CNTFET with the same configuration as that fabricated in [7] has been implemented. Figure 2 illustrates the simulated $I_D$-$V_{GS}$ characteristic of the simulated CNTFET (line) and compares it with the experimental result (stars) in [7]. Even though the experimental data are usually highly affected by parasitic resistance, which is difficult to correctly model in the absence of enough information about experimental structure, the comparison presents that the simulation results are in reasonable agreement with the existing experimental data. The operational temperature of the transistor usually differs from room temperature due to current drive and device resistances. To investigate the influence of temperature on the attributes of CNTFETs, the temperature has been changed in the conventional operating interval from 250 to 500 K. The output characteristic of the predefined structure at border temperatures of the interval is illustrated in Figure 3. It can be seen from the figure that at low gate source voltages, for higher temperature (500 K), the drain current is higher. In spite of this, by increasing the gate source voltage, at low drain source voltages, for higher temperature (500 K) the drain current is less than lower temperature (250 K). In the saturation region by increasing the gate source voltage the drain current difference between high and low temperature reduces. It is evident from the figure that the drain current in the saturation region and $V_{GS} = 0.8$ for 250 K and 500 K are approximately equal. Figure 4(a) shows on-state current versus temperature at $V_{GS} = 0.8$ V and $V_{DS} = 0.8$ V. The on-state current increases when temperature increases. The slope of this increase is approximately constant. Also, the temperature variation of off-current is illustrated in Figure 4(b) at $V_{GS} = 0$ V and $V_{DS} = 0.8$ V bias conditions. It is observed from the figure that the off-current in $T = 500$ K is 150 times higher than $T = 250$ K. Therefore, in spite of the on-current increase, the device reliability declines due to large rise in leakage current.

Figure 5(a) shows the transconductance characteristics of the CNTFET with the gate length of 20 nm for two different temperatures at $V_{DS} = 50$ mV and $V_{DS} = 0.8$ V. Although the on-state current increases when the temperature is increased from 250 to 500 K, the off-state leakage current grows considerably faster than that of the on-state current which confirms the previous results. Drain conductance ($g_d$) depends on the drain current in on-state. So, increase in on-current results in increase in drain conductance. The dependence of drain conductance on temperature at $V_{GS} = 0.8$ V and $V_{DS} = 0.8$ V is shown in Figure 5(b). It is evident from the figure that the drain conductance raises as the temperature increases by a factor of 2 from 250 to 500 K.

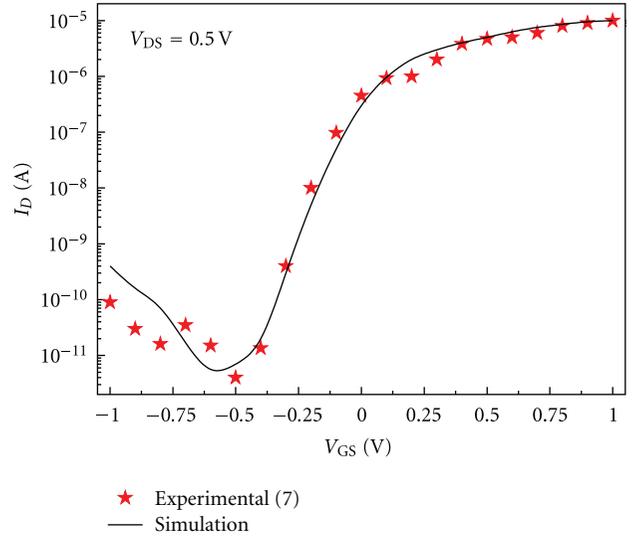

Figure 2: Comparison of the simulated $I_D$-$V_{GS}$ characteristic with that measured experimentally in [7] at $V_{DS} = 0.5$ V.

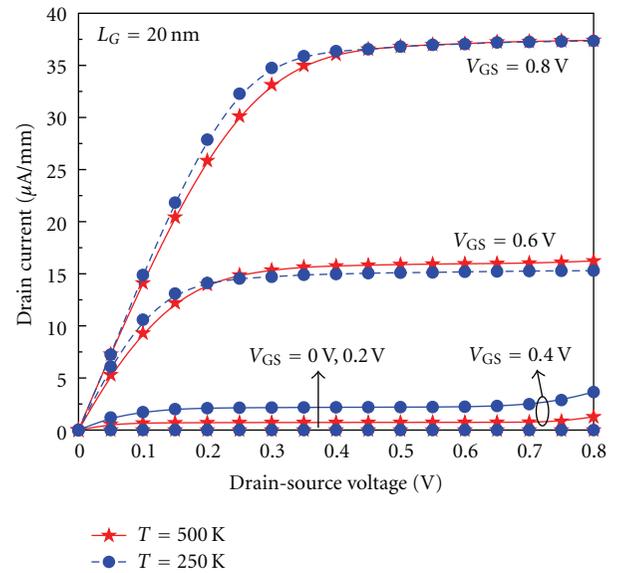

Figure 3: Comparison of the output characteristics ($I_D$-$V_{DS}$) of the CNTFET in 250 and 500 K for different gate biases.

Figure 6(a) illustrates the off-state leakage current versus saturation current ($I_{off}$–$I_{on}$) of CNTFET when temperature is varied from 250 to 500 K. The saturation current ($I_{on}$) is the drain current at $V_{GS} = 0.8$ V and $V_{DS} = 0.8$ V and the leakage current ($I_{off}$) is the drain current at $V_{GS} = 0$ V and $V_{DS} = 0.8$ V. It can be seen that the current capability of CNTFET degrades as the temperature is increased. Figure 6(b) shows the simulated on-/off-current ratio ($I_{on}/I_{off}$) of CNTFET for different temperatures. The on-/off-current ratio of CNTFET is significantly reduced in higher temperatures. By increasing the temperature the on-/off-current ratio decreases. It is apparent from the figure that the $I_{on}/I_{off}$ ratio of $1.1 \times 10^5$ at 250 K reduces to 650.4 at 500 K which shows the undesirable effects of increasing temperature in the current drive of the device. This ratio for $T = 500$ K is 169 times lower than $T = 250$ K. This decrease is apparent and more noticeable between 250 and 310 K. After 310 K the reduction slope decreases and after 410 K the on-/off-current ratio is approximately constant. The results imply that increasing the temperature declines the gate control on channel, and therefore the current capability. It can be concluded from Figures 6(a) and 6(b) that the increment in temperature not only increases the on-state current but also leads to a higher off-state leakage current.

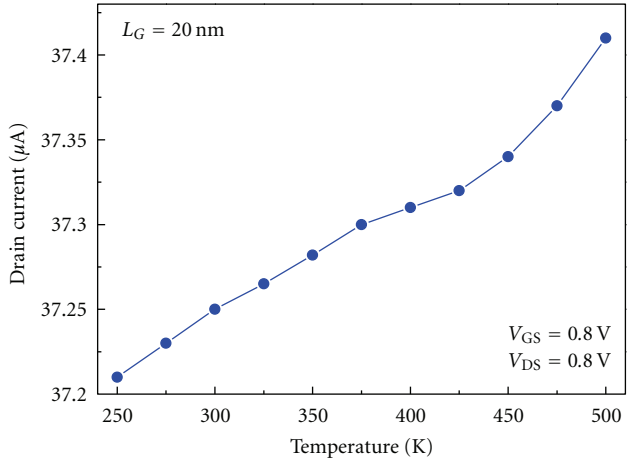

(a)

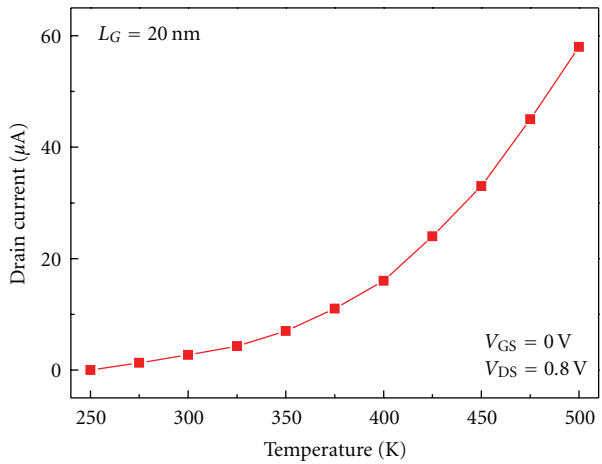

(b)

Figure 4: (a) On-state current versus temperature at $V_{GS} = 0.8$ V and $V_{DS} = 0.8$ V. (b) off-state current versus temperature at $V_{GS} = 0$ V and $V_{DS} = 0.8$ V.

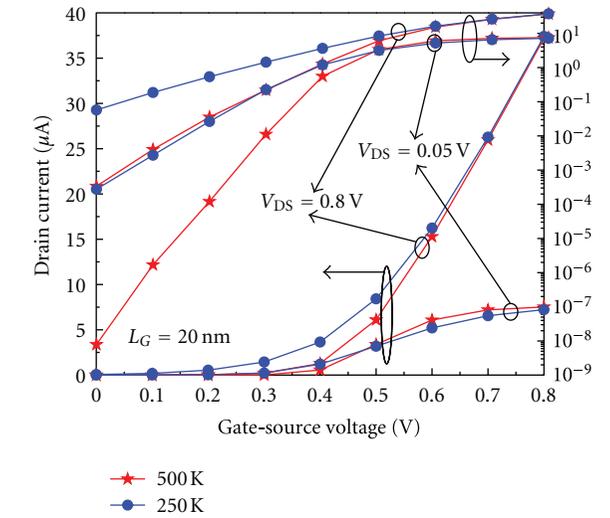

(a)

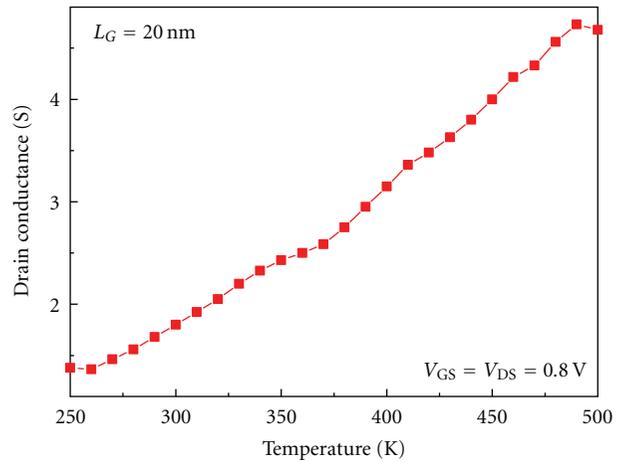

(b)

Figure 5: Variation of (a) transconductance ($g_m$) and (b) drain conductance ($g_d$) at $V_{DS} = 0.8$ V and $V_{GS} = 0.8$ V as a function of temperature.

The threshold voltage ($V_{th}$) is one of the most important parameters of the nanoscale devices. The variations in leakage and on-state currents by temperature indicate that the device threshold voltage changes by temperature. Therefore, it is necessary to investigate the influence of temperature variation on the threshold voltage characteristic of CNTFETs. Figure 7 shows the threshold voltage variation of the CNTFET structure as a function of the temperature. It can be seen that by varying the temperature, the device shows an approximately large $V_{th}$ variation which is undesirable for reliable CMOS applications.

It can be clearly observed that the slope of $V_{th}$ variation increases when the temperature increases from 250 to 500 K. The CNTFET structure exhibits lower threshold voltage roll-off at lower temperatures. This is more noticeable that for temperatures higher than 310 K the reduction slope increases with an approximately flat slope. This indicates that the roll-off of the threshold voltage is more severe for higher temperatures due to the short channel effects.

The subthreshold swing is a key parameter for transistor miniaturization. A small subthreshold swing is required to provide an adequate value of the on-/off-current ratio.

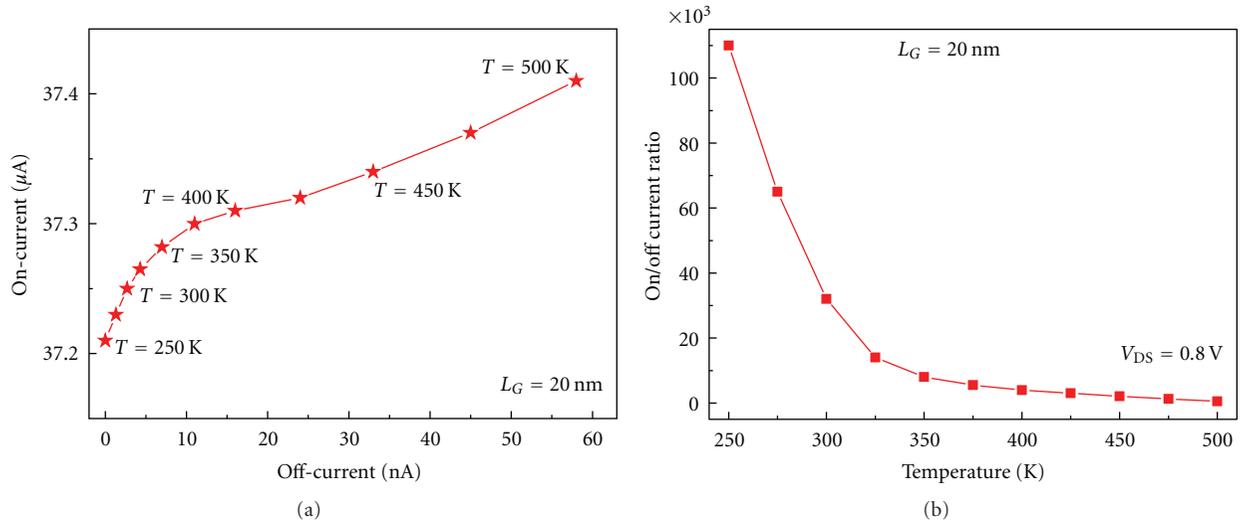

Figure 6: Temperature dependency of the (a) $I_{off}$-$I_{on}$ and (b) $I_{on}/I_{off}$ characteristics at $V_{DS} = 0.8$ V.

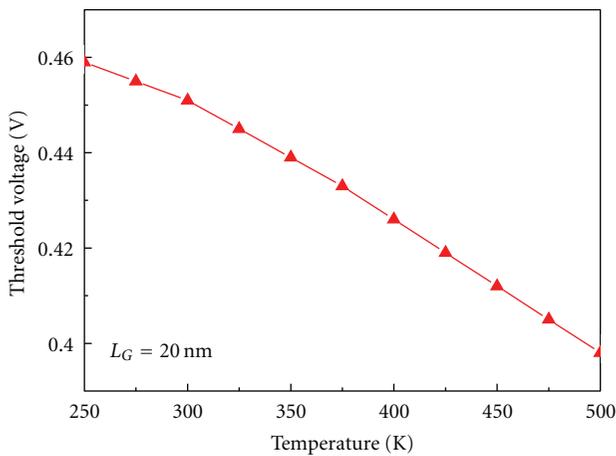

Figure 7: Threshold voltage variation characteristics as a function of temperature.

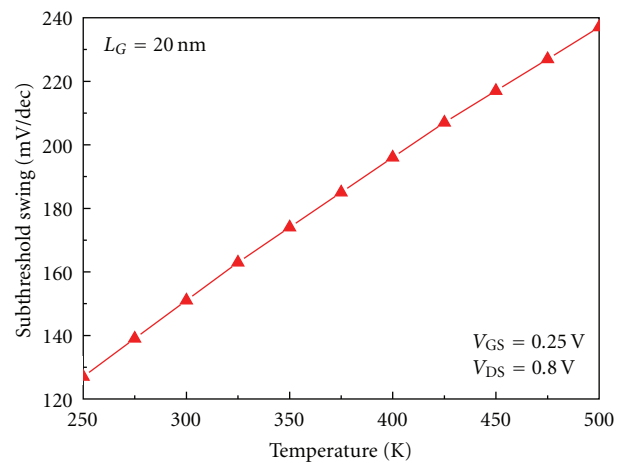

Figure 8: Subthreshold swing as a function of temperature.

Also, it is desired for low threshold voltage and low power operation FETs that scaled down to small sizes. The variation in subthreshold swing versus temperature for the CNTFET with 20 nm gate length is shown in Figure 8. It can be seen that the subthreshold swing increases with the increase in temperature. Approximately, the subthreshold swing increases by 4.35 mV in each temperature decade at $V_{GS} = 0.25$ V and $V_{DS} = 0.8$ V bias conditions.

As the channel length enters into the nanometer regime, many undesirable quantum and short-channel effects (SCEs) such as the threshold voltage roll-off and the DIBL become more apparent. These harmful effects cause deviation from the ideal performance of the CNTFETs. The DIBL effect is an electrostatic effect that can change the channel from a state of pinch-off to conduction and result in a substantial leakage current. It also shifts the threshold voltage and renders the gate ineffective in controlling the channel. Consequently, the DIBL effect degrades the device performance which should be avoided in device and circuit design. As can be seen from Figure 9 by increasing temperature the undesirable DIBL effect significantly increases. It should be noted that the DIBL effect occurs when the barrier height for channel carriers at the edge of the source reduces due to the influence of drain electric field, upon application of a high drain voltage. Similar mechanism can take place when the temperature increases. Therefore the energy of carriers increases and a large number of carriers injected into the channel, leading to an increased drain off-current. The relationship between the subthreshold swing and DIBL parameters of CNTFET from 250 K to 500 K is presented in Figure 10. It is interesting to notice that the subthreshold swing and DIBL parameters are almost linearly correlated, confirming that the degradation of these parameters has the same origin and can be perfectly influenced by the temperature.

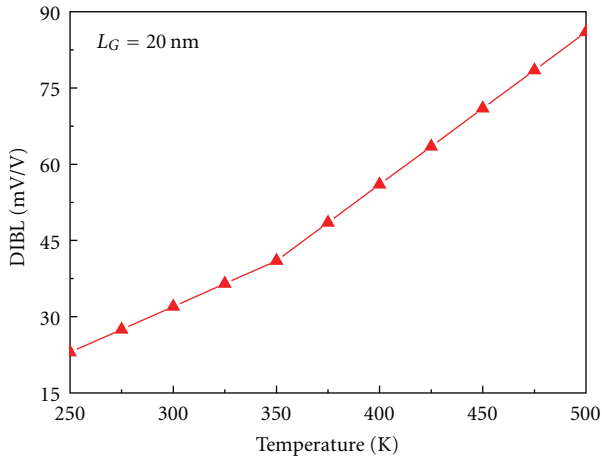

Figure 9: Drain-induced barrier lowering (DIBL) as a function of temperature.

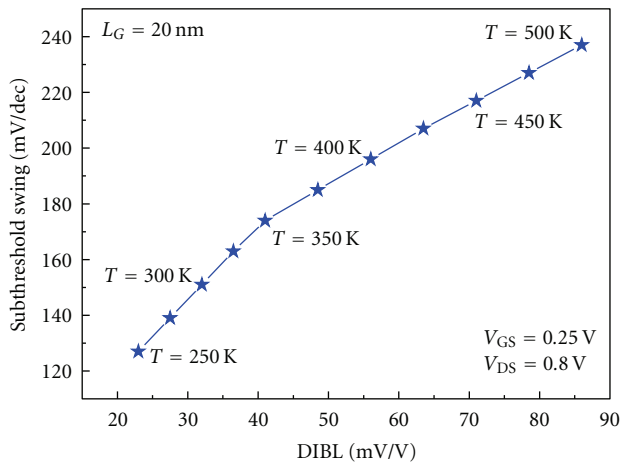

Figure 10: Relationship between the subthreshold swing and the DIBL parameters in different temperatures.

## 4. Conclusions

In this paper the attributes of carbon nanotube field-effect transistors (CNTFETs) by varying the temperature have been comprehensively investigated by developing a two-dimensional (2D) quantum simulation. By employing the nonequilibrium Green's function (NEGF) formalism and solving the Schrödinger and Poisson equations self-consistently, the influence of varying temperature on CNTFET performance in terms of on-current, off-current, on-/off-current ratio, transconductance, drain conductance characteristics, drain-induced barrier lowering (DIBL), threshold voltage, and subthreshold swing have been explained. The results show that the increase in temperature results in higher subthreshold swing and lower on-/off-current ratio. It is interesting to notice that the subthreshold swing and DIBL parameters are almost linearly correlated, confirming that the degradation of these parameters has the same origin and can be perfectly influenced by the temperature. These achievements can be effectively used for design considerations in these devices.

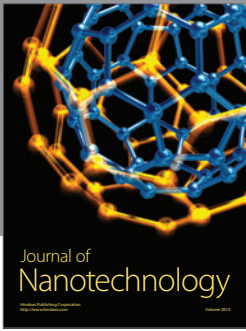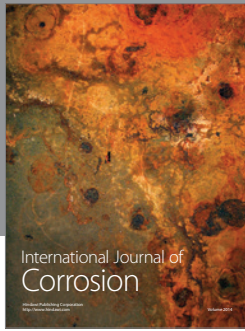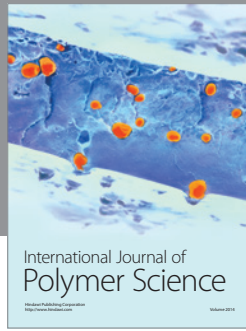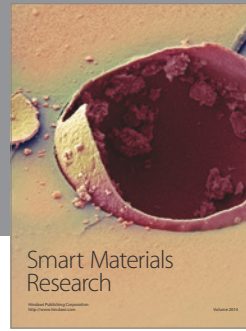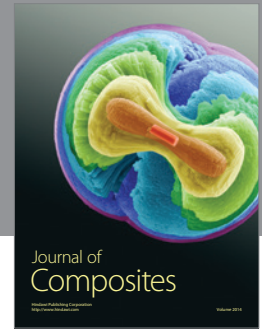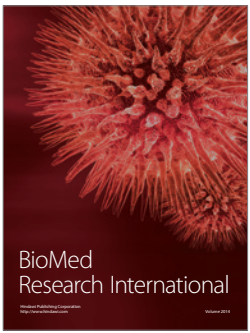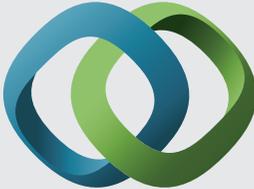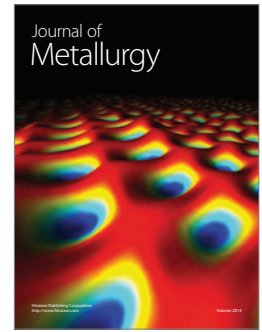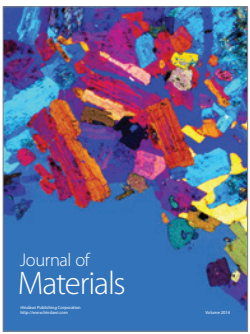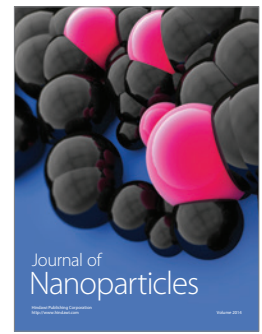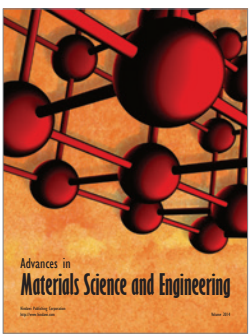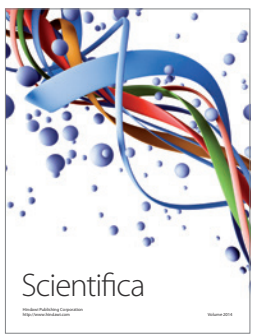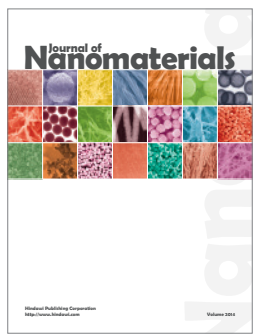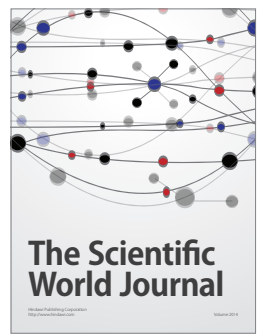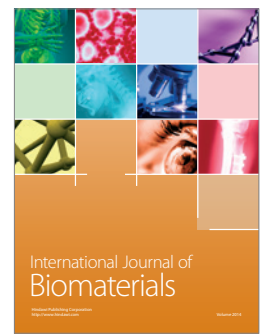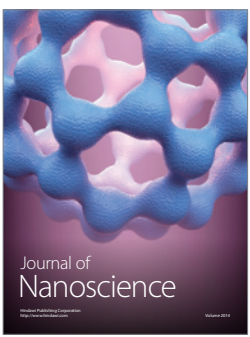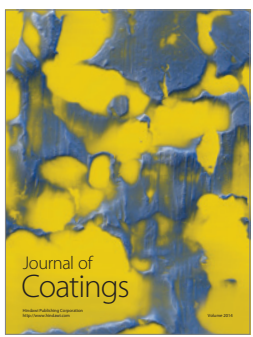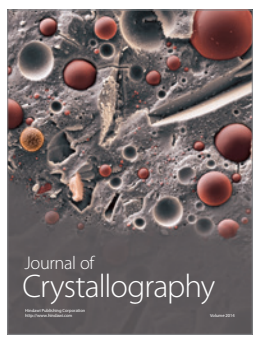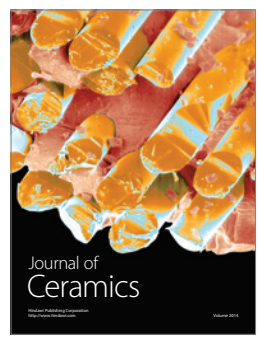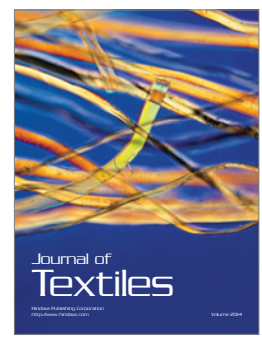